**Graphical Abstract**

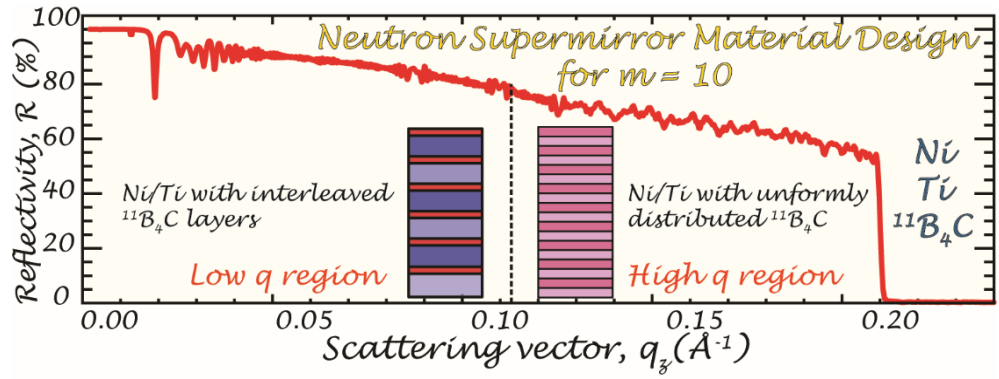

# Material design optimization for large-m $^{11}$B$_4$C-based Ni/Ti supermirror neutron optics

Sjoerd Stendahl,[a] Naureen Ghafoor,[a,*] Anton Zubayer,[a] Marcus Lorentzon,[a] Alexei Vorobiev,[b,c] Jens Birch,[a] Fredrik Eriksson [a]

[a]Department of Physics, Chemistry, and Biology, IFM, Linköping University, SE-581 83 Linköping, Sweden
[b]Department of Physics and Astronomy, Material Physics, Uppsala University, SE-751 20 Uppsala, Sweden
[c]Institut Max von Laue—Paul Langevin, 71 avenue des Martyrs, 38000 Grenoble, France
*Corresponding author: naureen.ghafoor@liu.se

**Abstract**

State-of-the-art Ni/Ti supermirror neutron optics have limited reflected intensity and a restricted neutron energy range due to the interface width. Incorporating low-neutron-absorbing $^{11}$B$_4$C enhances reflectivity and allows for thinner layers to be deposited, with which more efficient supermirrors with higher m-values can be realized. However, incorporating $^{11}$B$_4$C reduces the optical contrast, limiting the attainable reflectivity at low scattering vectors, making this approach infeasible. This study explores various approaches to optimize the material design of $^{11}$B$_4$C-containing Ni/Ti supermirrors to maintain high reflectivity at low scattering vectors and achieve low interface widths at large scattering vectors. The scattering length density contrast versus interface width is investigated for multilayer periods of 30 Å, 48 Å, and 84 Å, for designs involving pure Ni/Ti multilayers, multilayers with $^{11}$B$_4$C co-deposited in Ni and Ti layers, multilayers with $^{11}$B$_4$C co-deposited only in Ni layers, and multilayers with $^{11}$B$_4$C as thin interlayers between Ni and Ti layers. Our results suggest that a depth-graded hybrid material design by incorporating $^{11}$B$_4$C inside the Ni and Ti layers, below approximately 26 Å, and introducing 1.5 Å $^{11}$B$_4$C interlayers between the thicker Ni and Ti layers can achieve a higher reflectivity than state-of-the-art Ni/Ti multilayers over the entire scattering vector range.

## I. Introduction

Neutron scattering experiments have a major limitation due to the low flux of neutrons, which, despite improvements in modern neutron sources, remains several orders of magnitude lower than that of X-ray synchrotron sources [1]. However, advancements in optical components at neutron sources, such as supermirrors used in neutron guides, are expected to improve the neutron flux [2,3,4,5]. The reflectivity of supermirrors depends on the optical contrast between interfaces in terms of scattering length density (SLD) and interface width. The interface width also limits the minimum achievable layer thickness in the multilayer, limiting the attainable scattering vector q-range of the supermirror. The most commonly used method to account for the reduction of reflectivity caused by interface width is the Debye-Waller-like factor [6,7,8,9];

$$R = R_0 e^{-\left(2\pi n \frac{\sigma}{\Lambda}\right)^2}, \qquad (1)$$

where R is the resulting reflectance, $R_0$ is the theoretical reflectance without any interface width, n is the reflection order, σ the interface width, and Λ the period of the multilayer stack. Since this expression contains the ratio of the interface width to the multilayer period, this factor shows that the interface width becomes increasingly important for thinner periods, while thicker layers exhibit a weaker dependence on interface width. This is particularly relevant for supermirrors, where a depth-graded layer thickness design incorporates both thick and thin layers in the multilayer stack [10,11].

Typically, the Ni/Ti material system is used for supermirrors, providing excellent optical contrast for maximizing reflectivity. The interface width in these multilayers is mainly attributed to nanocrystallites, intermetallic formation, intermixing, and interdiffusion between adjacent layers in the multilayer. Since the reflectivity performance of multilayered structures strongly depends on achieving a small interface width, this has been widely researched in the past. Many different techniques have been studied. One commonly used method is to grow thin interlayers (<0.3 nm) of Cr [12] or B$_4$C [13,14] in between the interfaces to inhibit diffusion, which can result in more abrupt interfaces in the multilayer stack. This reduces the interface width and improves the overall performance of the multilayer. Additionally, the asymmetry in surface free energy between Ni and Ti can lead to rougher interfaces and stresses in the multilayer. To address this issue, growing an intermediate layer of Ag or B$_4$C can reduce this asymmetry [14], further improving the Finterface quality [15]. It has been reported that the thickness of the Ni crystalline transition is approximately 2 nm [16].

Several studies have shown that adding C atoms can inhibit Ni crystallization and suppress interdiffusion between layers in NiC/Ti multilayers [17,18,19]. Furthermore, our group has previously demonstrated that by co-depositing $B_4C$ throughout Cr/Sc X-ray multilayers, the multilayers can be amorphized, eliminating nanocrystallites at the interfaces. [20].

Incorporating low-neutron-absorbing $^{11}B_4C$ during growth can eliminate nanocrystallite formation by amorphizing the layers, and prevent intermetallic formation and interdiffusion at interfaces, as demonstrated in previous research [2,3,21]. Our earlier work showed that adding $^{11}B_4C$ into Ni/Ti multilayers with a period of 48 Å significantly improves reflectivity performance by reducing interface widths [2,3,22]. We significantly improved the reflectivity of Ni/Ti-based multilayers and achieved interface widths of 4.5 Å by combining $^{11}B_4C$ co-sputtering with a modulated ion-assisted deposition scheme [3]. These findings demonstrate the great potential of this deposition technique for application in neutron optical instruments.

Reflectivity simulations using an interface width of 4.5 Å show remarkable improvement near the critical edge of m = 6 supermirrors, where m is defined as the critical angle of the supermirror divided by the critical angle of Ni in bulk. However, at lower incidence angles, the predicted reflectivity performance was still lower than that of commercially available Ni/Ti supermirrors with an interface width of 7 Å [2,23], as demonstrated in Figure 1. This lower reflectivity is attributed to the dilution of optical contrast in terms of neutron scattering length density (SLD) due to the incorporation of $^{11}B_4C$ into all layers of the multilayer stack. Since the lower interface width achieved by adding $^{11}B_4C$ is less important for thicker layers (Eq. 1), it is not sufficient to compensate for the loss in optical contrast by reducing the interface width further.

This trade-off between a large contrast in SLD and a low interface width gives rise to two distinct regions, where a pure Ni/Ti multilayer with an interface width of σ = 7 Å performs better at reflections below a scattering vector of $q_z$ = 0.11 Å$^{-1}$, corresponding to a multilayer period of Λ = 58 Å, while a $^{11}B_4C$-containing multilayer with an interface width of σ = 4.5 Å has a higher reflectivity above this transition. The ideal material design for a neutron multilayer, therefore, depends on the relevant scattering vector q-range, where achieving small interface widths is crucial at high q-values, while a large SLD contrast is more important at low q-values, corresponding to thicker periods. It should be noted that this transition region depends on the interface width and the amount of $^{11}B_4C$ incorporated in the supermirror.

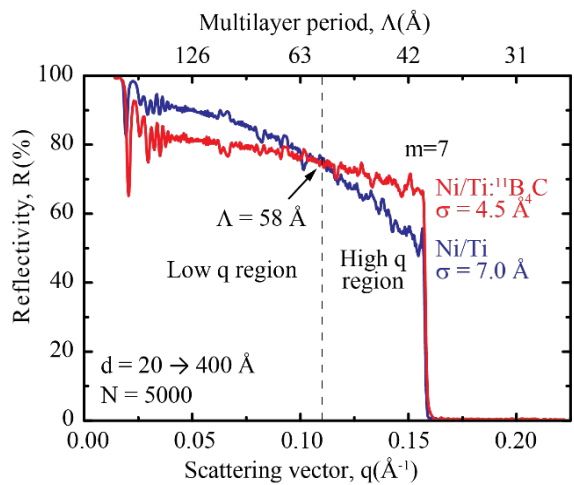

**Figure 1.** a) Simulated reflectivity performance of state-of-the-art Ni/Ti and $^{11}B_4C$-containing Ni/Ti m = 7 supermirrors, with interface widths of 7 Å (blue) and 4.5 Å (red), respectively. The dashed line indicates the transition point where a large contrast in SLD is most important for reflectivity, and where a low interface width becomes crucial.

Despite previous research efforts, there has been a lack of consideration for varying the materials design with multilayer periodicity to achieve the ideal supermirror. In this work, several approaches for the optimization of neutron multilayers in different regions of q-space are presented, addressing the lower predicted reflectivity at low scattering vectors for the $^{11}B_4C$-containing supermirrors presented in our earlier work. Here, we divide this into two distinct regions in q-space, where the low-q region represents multilayers with a thick period where the optical contrast is most important, while the high q-region represents multilayers with a thinner period where the interface width needs to be maximized to obtain the highest reflectivity. We aim to address the low reflectivity at low scattering vectors for the $^{11}B_4C$-containing supermirrors presented in our earlier work.

Figure 2 illustrates the various multilayer designs and corresponding simulated SLD profiles investigated in this study. The first design, in Figure 2 a), shows a conventional Ni/Ti multilayer. This design achieves good reflectivity at low q-values due to the high contrast in SLD, but limited reflectivity at high q-values due to the higher interface width. The second design shown in Figure 2 b), involves co-deposition with a continuous flux of boron and carbon with Ni and Ti during the entire multilayer growth using co-sputtering with an $^{11}B_4C$ target material. This technique reduces the interface width, but limits reflectivity at low q-values due to reduced optical contrast (see. Fig. 1 and Ref. [25]. The dilution in optical contrast is shown by the SLD profile (black dotted line) in Fig. 2b). The third design, in Figure 2 c), shows a multilayer where $^{11}B_4C$ is co-deposited into the Ni layer only, to reduce the nanocrystallite formation which predominantly forms in the Ni layers similar to the NiC/Ti multilayers [17,18,19]. The close match in neutron SLD between Ni and $^{11}B_4C$, $9.4 \cdot 10^{-6}$ Å$^{-2}$ and $9.1 \cdot 10^{-6}$ Å$^{-2}$, respectively [24], preserves the high optical contrast of the pure multilayer. However, the missing $^{11}B_4C$ in the Ti layer can cause it to crystallize, resulting in rougher interfaces and a higher interface width. The fourth multilayer design, as shown in Figure 2 d), comprises of 1.5 Å thin $^{11}B_4C$ layers that are deposited at the interfaces of a Ni/Ti multilayer. The purpose of using these layers is to prevent interdiffusion and intermetallic formation, which ultimately reduces total interface width and enhances reflectivity.

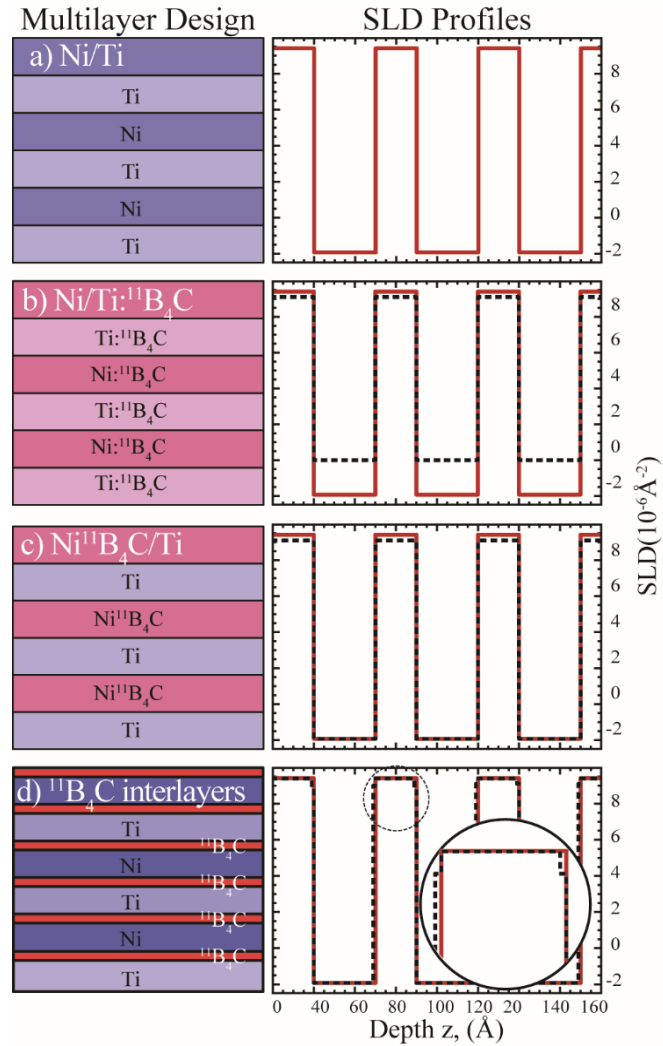

Figure 2: Multilayer designs and simulated SLD profiles investigated in this study. a) Pure Ni/Ti multilayers, b), Ni/Ti:$^{11}B_4C$ multilayers where both Ni and Ti layers are co-deposited with $^{11}B_4C$, c) Ni$^{11}B_4C$/Ti multilayers where $^{11}B_4C$ is only co-deposited in the Ni layer, and d) Ni/Ti multilayers with $^{11}B_4C$ interlayers at each interface. SLD profiles of designs b), c), and d) (black solid lines) are compared with the SLD profile in a) of a pure Ni/Ti multilayer (red dashed line). In d) the step in SLD at the interface is due to $^{11}B_4C$ and is magnified in the inset.

The four multilayer designs can be categorized into two distinct approaches. The Ni$^{11}B_4C$/Ti and the multilayers with $^{11}B_4C$ interlayers maintain the high optical contrast present in pure Ni/Ti and are, therefore, expected to perform well for thicker periods. The Ni/Ti:$^{11}B_4C$ multilayers, on the other hand, primarily focus on reducing the interface width and are expected to perform well for thinner periods. Where Ni/Ti:$^{11}B_4C$ is not studied for the thicker periods, the two multilayer designs Ni$^{11}B_4C$/Ti multilayers and the multilayers with $^{11}B_4C$ interlayers are completely new approaches for the entire q range tested in this work.

A multitude of analytical techniques were utilized to better comprehend how various growth parameters and multilayer material designs affect the structural and optical properties of the multilayers. These included X-ray and neutron reflectivity, X-ray diffraction, grazing-incidence small-angle X-ray scattering, and transmission electron microscopy. The obtained data was used to optimize multilayer designs for different periodicities, to improve the reflectivity at low scattering vectors, which was previously limited in $^{11}B_4C$-containing multilayers. Combining the optimal design parameters, a hybrid multilayer design was predicted to outperform the current state-of-the-art, based on its improved reflectivity performance.

## II. Experimental details

*Multilayer deposition*

All multilayers investigated in this study were deposited using triple cathode direct current magnetron sputtering in a high vacuum system onto 10×10×0.5 mm³ Si (001) substrates with a native oxide. The background pressure before deposition was approximately $6.7·10^{-6}$ Pa ($5.0·10^{-7}$ Torr). A pressure of 3 mTorr Ar (99.999% purity) was used as sputtering gas and ambient temperatures (293 K) were used during the depositions. The substrate was rotating at a constant speed of 17 rpm during deposition.

The three sputtering targets, Ni (99.99% purity, ⌀75 mm diameter), Ti (99.95%, ⌀75 mm), and $^{11}B_4C$ (98.7%, ⌀50 mm), were continuously running during deposition and their respective material fluxes were controlled by using computer-controlled shutters in front of the magnetrons. Ni and Ti discharges were established with current-regulated power supplies, with 80 mA and 160 mA, respectively, while the $^{11}B_4C$ magnetron was operated using power regulation at 30 W.

A negative substrate bias voltage was applied to the substrate table to attract ions from the sputtering plasma to the growing multilayer. A high flux, two-phase, modulated ion assistance scheme was employed [25], where the initial 3 Å layer was deposited with a grounded substrate bias, followed by the growth of the remaining layer with a higher substrate bias of -30 V. The high flux of ions was generated using a magnetic coil surrounding the substrate, which magnetically guided secondary electrons to the substrate region, thereby increasing the plasma density near the substrate surface and enhancing the ion-to-adatom ratio. This approach has been shown to reduce roughness and eliminate intermixing in multilayer systems. Further details about the deposition system and the ion assistance design can be found elsewhere [2,3].

In this study, various multilayer designs (see Figure 2) were grown with different periods of 30 Å, 48 Å, and 84 Å. These periods were chosen to study multilayers with first-order Bragg peaks distributed in the different scattering vector regions illustrated in Figure 1. Multilayers were grown with periodicities of $\Lambda$ = 48 Å and N = 50 periods, as well as with periodicities of $\Lambda$ = 30 Å and N = 80 periods, resulting in a nominal total multilayer thickness of 240 nm. Additionally, a stacked design was deposited, consisting of three multilayers with different periods deposited on each other. The bottom multilayer had a period of $\Lambda$ = 30 Å and N = 50 periods, the middle multilayer had a period of $\Lambda$ = 48 Å and 19 periods, and the top multilayer had a period of $\Lambda$ = 84 Å and N = 33 periods. Each multilayer within the stack had a nominal thickness of 158 nm, resulting in an equal total thickness for each period. For the multilayers with thin interlayers, an interlayer thickness of 1.5 Å was used, with the thickness of the Ni layer reduced to maintain the same effective thickness ratio. For the GISAXS measurements, multilayers were grown at a period of $\Lambda$ = 48 Å, and a larger number of periods at N = 100, in order to increase the statistics in the measured scattering signal.

Furthermore, one multilayer stack was grown where the top multilayer with a period of 84 Å consisted of pure Ni/Ti, while the middle and bottom multilayers with periods of 48 Å and 30 Å, respectively, consisted of Ni/Ti:$^{11}B_4C$ with $^{11}B_4C$ incorporated throughout these periods. This design demonstrates how a depth-graded multilayer can be grown with different material systems depending on the multilayer period. All multilayers were deposited under the same deposition conditions, with a high flux and modulated low energy ion assistance during growth.

*Multilayer characterization*

The elemental composition of the films was determined using time-of-flight energy recoil detection analysis (ToF-ERDA). A primary beam of $^{127}I^{8+}$ was used with an energy of 36 MeV at an incident angle of 67.5° relative to the surface normal, with the energy detector placed at a recoil scattering angle of 45°. A detailed description of the experimental set-up is available elsewhere [26,27]. The measured data was analyzed using the Potku software, [28] where the measured recoil energy spectrum of each element was converted to relative atomic concentrations with an accuracy of ±0.5 at.%. Measurements of the Ni/Ti:$^{11}B_4C$ multilayer reveal 16.2 at.% $^{11}B$ and 3.6 at.% C. Based on the deposition timings for each multilayer, an estimate for the individual layer content has resulted in approximately 20.5 at.% and 4.6 at.% C, in the Ti layers, and about 11.9 at.% $^{11}B$ and 2.6 at.% C in the Ni layers.

GISAXS measurements were performed at the PETRA III synchrotron in Hamburg, at the Microfocus small- and wide-angle X-ray scattering P03 beamline [29]. A monochromatic X-ray beam of 0.96 Å wavelength at a fixed incidence angle of 0.4° was used to ensure a high intensity and sufficient penetration depth to resolve the entire multilayer stack [30]. The sample-to-detector distance was 3850 mm, and a PILATUS 2M detector system was used to collect 2D intensity GISAXS maps. In-house developed software for data reduction and analysis [31] was used to obtain GISAXS line scans by performing line integrations over selected areas of interest. Structural information about the lateral interface morphology of each sample was obtained by determining the full width at half maximum (FWHM) of the Bragg sheet in the in-plane direction. In addition, the FWHM of the Bragg sheet in the out-of-plane $q_z$-direction was determined at various in-plane $q_y$-positions to investigate the dependence of the vertical correlation length on the spatial frequency in $q_y$ [21, 32]. The vertical correlation is commonly quantified by the number of effectively correlated periods, which is inversely proportional to the FWHM of the Bragg sheet in the growth direction and can be expressed as $N_{eff} = 2/(FWHM(q_y) \cdot \Lambda)$ [33]. FWHM values at positive and negative $q_y$-values were obtained and averaged for better statistical analysis.

X-ray diffraction (XRD) was performed using a Panalytical X'Pert Bragg-Brentano θ-θ diffractometer with Cu-K$_\alpha$ X-rays. On the primary side, a Bragg-Brentano HD mirror was used with a ½° divergence slit and a ½° anti-scatter slit, and on the secondary side a 5 mm anti-scatter slit was used together with an X'celerator detector operating in scanning line mode. Diffraction measurements were performed in the range 20°-80° 2θ with a step size of 0.033°/step and a time per step of 50 s, leading to a total acquisition time of approximately 12 min.

Neutron reflectometry measurements were conducted for a selection of samples at the Institut Laue-Langevin in Grenoble using the Swedish neutron reflectometer SuperADAM [34]. The measurements were performed using a monochromatic wavelength of 5.23 Å and a sample-to-detector distance of 150 cm. To account for the significantly higher measured intensity at lower incidence angles, the measurements of the single multilayers were divided into different regimes, each with higher acquisition times at higher angles. This resulted in a total acquisition time of approximately 5 hours per sample. The stacked multilayers were all measured in a single measurement with a total acquisition time of 3.3 hours per sample. Footprint correction was applied to all samples using dedicated data reduction software used for the correction. The data set was then normalized to the obtained critical angle.

X-ray reflectivity measurements were performed using a Panalytical Empyrean diffractometer with a Cu X-ray tube. A parallel beam X-ray mirror was used on the primary beam side in combination with a 1/32° divergence slit, while a parallel plate collimator was used in combination with a collimator slit on the diffracted beam side with a PIXcel3D detector in 0D mode. The reflectivity of the multilayers was measured in the range 0°-10° 2θ with a step size of 0.01°/step and a collection time of 0.88 s per step, giving a total measurement time of approximately 30 min.

Structural parameters such as multilayer period, layer thickness ratio, and interface widths, were obtained by fitting the X-ray and neutron reflectivity data simultaneously to a model created within the GenX reflectivity fitting software [35]. In this model the structural parameters are coupled, giving a single fit to two independent data sets for increased reliability. More details can be found elsewhere [3]. In this study, the used optical properties in terms of scattering length density were determined based on an interpolation between the tabulated bulk values of the constituent materials. The accuracy of these calculated values was confirmed by comparing the simulated critical angle with the experimental X-ray and neutron reflectivity data. To model the $^{11}B_4C$ interlayers, these were treated as pure Ni/Ti multilayers with the interlayers being a part of the interface width.

For TEM imaging, cross-sectional specimens were prepared by conventional mechanical polishing followed by Ar ion etching at 5 keV. Finally, the specimens were subject to Ar ion etching at 2 keV to remove surface damage resulting from the etching. The TEM investigations were performed using an FEI Tecnai G2 TF 20 UT field-emission TEM operated at 200 keV for a point resolution of 0.19 nm.

## III. Results and Discussion

X-ray diffraction (XRD) measurements were performed on multilayers with varying designs and multilayer periods, and the results are presented in Figure 3. In single multilayers with a period of 30 Å, the only diffraction peak observed was the Si 004 reflection at a 2θ position of approximately 69.1°, originating from the single-crystalline Si substrate. Additionally, a faint and broad low-intensity hump was detected in the range of approximately 35°-50° 2θ for all designs regardless of $^{11}B_4C$ incorporation, indicative of an X-ray amorphous multilayer structure. However, previous observations using high-resolution transmission electron microscopy have revealed that crystallites form in Ni/Ti multilayers with periods as thin as 16 Å, whereas the corresponding $^{11}B_4C$-containing ones are amorphous when observed using both X-rays and electrons [3].

For multilayers with a thicker period of 48 Å, the XRD patterns for $^{11}B_4C$-containing Ni/Ti and Ni$^{11}B_4C$/Ti remained identical, indicating that the layers are at least X-ray amorphous. Ti layers are intrinsically X-ray amorphous at these periods irrespective of $^{11}B_4C$ incorporation. In contrast, a broad diffraction peak appeared around 2θ = 44.5° for pure Ni/Ti multilayers and those with $^{11}B_4C$ interlayers, corresponding to diffraction from Ni 111 crystallites. The width of the peaks, which is associated with the crystallite size in the growth direction, indicates that the crystallites are extremely small and likely limited to the layer thickness, as is often observed in these types of metallic multilayers. The slightly lower peak intensity for multilayers with $^{11}B_4C$ interlayers than pure Ni/Ti multilayers suggests that more crystallites are present without interlayers. These crystallites, which exhibit a strong Ni 111 texture, result in facets at the interfaces, leading to roughness in the growth direction.

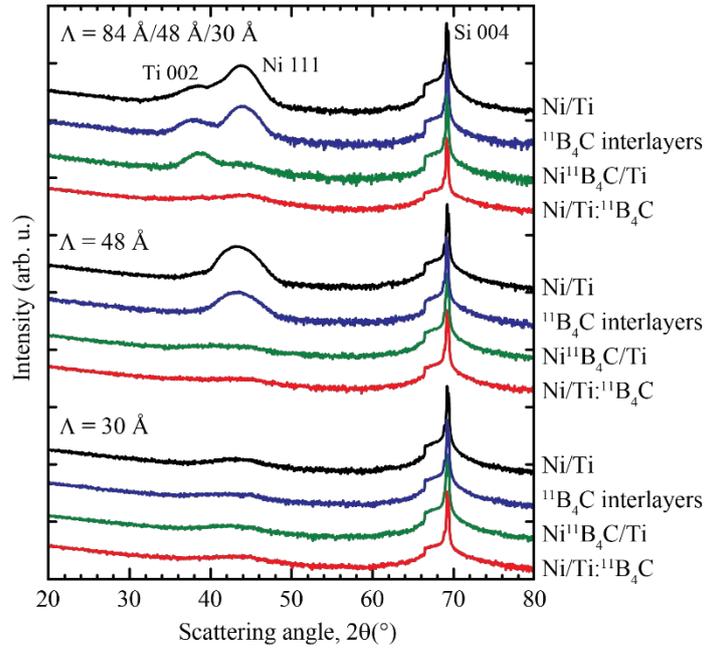

**Figure 3.** The obtained XRD results for the investigated material systems grown at different periods. The top curves are multilayers where three different periodicities are present in the same stack.

XRD measurements were also performed on the various designs of stacked multilayers consisting of three multilayers with periods of 30 Å, 48 Å, and 84 Å. In these measurements, an additional diffraction peak was observed at a diffraction angle of $2\theta = 38.4°$ for all multilayer stacks, except the Ni/Ti:$^{11}B_4C$, corresponding to diffraction from Ti 002 lattice planes, confirming the presence of Ti crystallites in the thicker layers of these multilayers. Notably, Ni$^{11}B_4C$/Ti exhibited a more prominent diffraction peak for Ti than for Ni, indicating that the addition of $^{11}B_4C$ to the Ni layers effectively suppresses Ni crystallite formation, as reported previously for NiC/Ti multilayers [17,18,19].

Furthermore, the Ni 111 diffraction peaks observed in the XRD patterns were narrower for the 84 Å period samples compared to those for the 48 Å period samples, indicating that larger crystallite sizes are allowed to form in the thicker layers which can lead to higher interface roughness. The higher intensity of the Ni 111 peak suggests more crystallites in the Ni/Ti multilayers.

Transmission electron microscopy (TEM) investigated three different multilayer stacks. Bright-field TEM micrographs of the multilayers where $^{11}B_4C$ was co-deposited throughout the entire stack are shown in Figure 4 a), while Figure 4 b) shows a hybrid stack where the thinner periods of 30 Å and 48 Å were co-deposited with $^{11}B_4C$, while the thicker period of 84 Å was grown using pure Ni/Ti. Figure 4 c) displays a Ni/Ti multilayer stack where $^{11}B_4C$ interlayers were deposited between the layers. The insets in Figure 4 a)-c) show the selected area electron diffraction (SAED) patterns of the multilayer diffraction spots acquired from the center of each multilayer, recorded with a beam parallel to the [110] zone axis of the Si substrate. The bottom two multilayers with periods of 30 Å and 48 Å, in the multilayer stacks in Figure 4 a) and b) were deposited with identical design parameters, reflected by the indistinguishable appearance of the micrographs. Both stacks exhibit high-quality multilayers with smooth interfaces for these periodicities. Recently, high neutron reflectivity has been reported from a well-defined 104-period Ni/Ti:$^{11}B_4C$ multilayer with 48 Å with 4.5 Å interface width [3]. The 30 Å period multilayers in this study demonstrate that the incorporation of $^{11}B_4C$ generates well-defined layers with smoother interfaces through stabilized amorphization also for the thinner periods. The 30 Å period multilayer in the multilayer stack in Figure 3 c), with interlayers, also exhibits amorphous layers; however, with less defined interfaces.

The threshold thickness for Ni layers to become amorphous to minimize the interface energy contribution is less than 1 nm. Although $^{11}B_4C$ interlayers facilitate amorphization for 30 Å periods, they do not fully suppress the formation of nanocrystallites and their associated interface width. This effect is more pronounced for the 48 Å multilayer (Figure 4 c), where bright and dark contrast, primarily in the Ni layers, indicates that the interlayers do not suppress crystallization in the bulk of the layers for periods larger than or equal to 48 Å. However, the $^{11}B_4C$ interlayers hinder intermetallic phase formation and amorphize the interface region, resulting in abrupt and smooth interfaces with reduced crystallite sizes for both 48 Å and 84 Å period multilayers. The amorphization effect of $^{11}B_4C$ is also evident in the multilayer stack shown in Figure 4 a) for 84 Å period compared to the corresponding periods in the multilayers in 4 b) and c), which clearly shows nanocrystallites, predominantly in Ni layers, where the individual Ni and Ti layers thicknesses constrain the vertical size. Additionally, the comparison of the multilayer reflections in the inset SAED patterns shows that higher order reflections are present in the Ni/Ti:$^{11}B_4C$ and the hybrid designs as compared to stack with $^{11}B_4C$ interlayers, indicating more abrupt interfaces when $^{11}B_4C$ is incorporated throughout the multilayer stack.

The effects described above are also evident in the dark-field TEM micrographs presented in Figure 4 d) and e), which were recorded using the Ni 111 diffraction spot in. For 30 Å periods, no observable difference in amorphous structure is apparent between the Ni/Ti:$^{11}B_4C$ multilayers and those with $^{11}B_4C$ interlayers in Figure 4 d) and 4 e), respectively. In contrast, for the 48 Å period Ni/Ti multilayer with $^{11}B_4C$ interlayers a strong diffraction contrast is observed. Moreover, comparing Ni/Ti with and without $^{11}B_4C$ interlayers for 84 Å periods shows that the interfaces are more abrupt with interlayers.

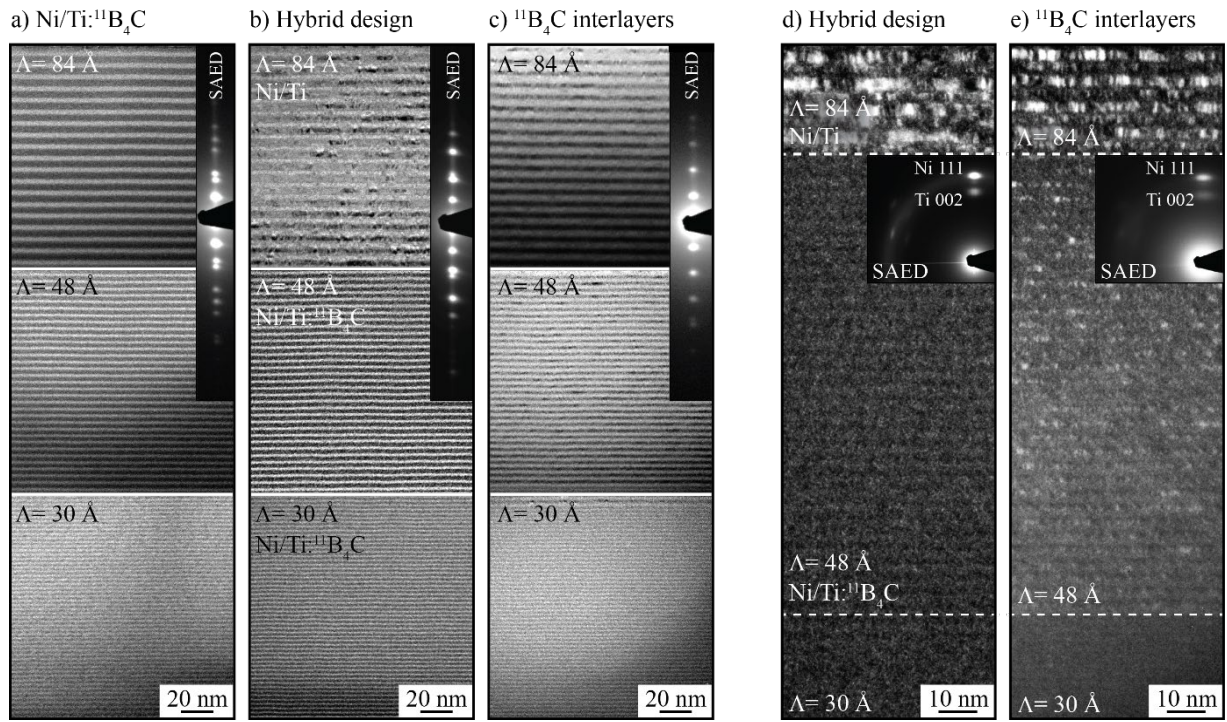

**Figure 4.** Bright field TEM micrographs of multilayers with periods of 30 Å, 48 Å, and 84 Å from three different multilayer stacks: a) Ni/Ti:$^{11}B_4C$, b) hybrid design with Ni/Ti:$^{11}B_4C$ in the bottom two multilayers and pure Ni/Ti in the top one, and c) Ni/Ti with $^{11}B_4C$ interlayers. The insets in the micrographs are SAED patterns showing multilayer reflections aligned with [110] zone axis of the substrate. Dark field TEM images of the stacked multilayers shown in b) and c) are presented in d) and e), respectively. The insets in d) and e) show the SAED patterns of the entire stack.

The SAED patterns in Figure 4 d) and e) show distinct diffraction spots corresponding to Ni 111 and Ti 002 lattice planes, originating mainly from the top 84 Å period multilayer, and a faint intensity ring resulting from the amorphous regions in the two bottom multilayers with periods 30 Å and 48 Å. Faint reflections are also observed for the pure Ni/Ti multilayer in 4 a), which are likely associated with a Ni-Ti intermetallic phase, although this is not conclusively determined in this study. The well-defined diffraction spots, with the strongest Ni 111 and Ti 002 reflections in the growth direction, suggest that the Ni and Ti layers are 111 and 002 textured, respectively, as expected for face-centered cubic and hexagonally close-packed metals due to surface energy minimization during growth.

Furthermore, it can be observed that the diffraction spots in Figure 4 d) are more well-defined laterally, indicating a slightly higher degree of texture for the pure Ni/Ti multilayers, while in the growth direction, the diffraction spots are narrower for the Ni/Ti multilayers with $^{11}B_4C$ interlayers in 4 e), indicating smaller crystallite sizes for these layers. This suggests that the $^{11}B_4C$ interlayers have a positive effect on reducing crystallite faceting at the interfaces.

While a TEM study was not performed for the Ni$^{11}B_4C$/Ti multilayer stack, it can be inferred from the combined information obtained in Figure 4 that Ni$^{11}B_4C$/Ti multilayers with 30 Å and 48 Å periods are likely to have smooth interfaces and amorphous layers except for indiscernible crystallites in the Ti layers, similar to what is shown in Figure 4 c) for the Ni/Ti with $^{11}B_4C$ interlayers. For a period of 84 Å amorphous Ni$^{11}B_4C$ layers can be expected, similar to the top multilayer shown in Figure 4 a), while the Ti layers are likely to be nanocrystalline, similar to the Ti layers shown for the top multilayer in the hybrid design in Figure 4 b).

Using grazing incidence small angle X-ray scattering (GISAXS), the effect of adding $^{11}B_4C$ interlayers to pure Ni/Ti multilayers and Ni/Ti:$^{11}B_4C$ has been compared. Figure 5 a) shows the results of GISAXS line scans across the Bragg sheet in the lateral direction for Ni/Ti, Ni/Ti:$^{11}B_4C$, and Ni/Ti with $^{11}B_4C$ interlayers. The intensity profiles reveal that the pure Ni/Ti multilayer has a different interface morphology than those containing $^{11}B_4C$, where the two latter display clear shoulders, indicating the formation of mounded interfaces. In previous studies, it was found that the addition of $^{11}B_4C$ to Ni/Ti multilayers resulted in mounded interfaces, while pure Ni/Ti multilayers exhibited characteristics typical of self-affine interfaces [21]. Here, the mounded interfaces are also confirmed for Ni/Ti multilayers with $^{11}B_4C$ interlayers. By estimating the characteristic length of the mound separation from the intersection between the tangents on either end of the shoulders in Figure 5 a) on a log-log scale, the resulting characteristic length between the mounds was found to be 410 Å and 483 Å for Ni/Ti:$^{11}B_4C$ and $^{11}B_4C$ interlayers, respectively. This indicates that the interface mounds in the Ni/Ti:$^{11}B_4C$ have a closer average separation, i.e. a higher density of mounds.

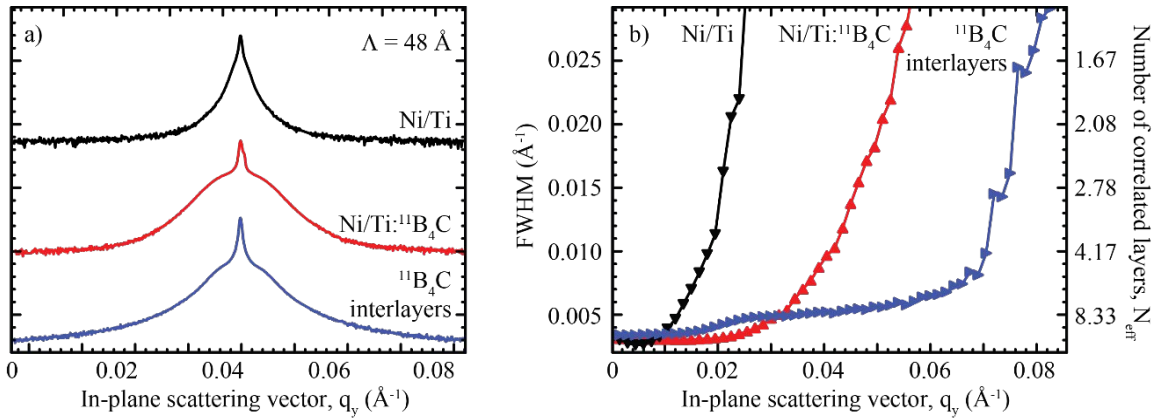

**Figure 5.** a) GISAXS line scans of the first Bragg sheet for Ni/Ti, Ni/Ti:$^{11}B_4C$, and Ni/Ti multilayers with $^{11}B_4C$ interlayers, all with a multilayer period of 48 Å. The interface profiles reveal a mounded interface morphology when $^{11}B_4C$ is present. b) Variation of the FWHM of the Bragg sheet in the growth direction and the corresponding effective number of correlated periods as a function of the in-plane scattering vector. The results demonstrate that incorporating $^{11}B_4C$ leads to stronger correlations at higher lateral spatial frequencies.

Figure 5 b) shows the FWHM of GISAXS line scans along the $q_z$-direction at the first Bragg sheet at different positions in $q_y$ for the multilayer designs that were investigated. These scans provide information about the effective number of correlated layers at each lateral $q_y$-position. Consistent with previous reports, pure Ni/Ti multilayer exhibit strong correlation at low $q_y$-values, while the correlation decreases rapidly at higher $q_y$-values, indicating that the large-scale features at the interfaces are repeated, but smaller details corresponding to lower lateral frequencies are not replicated throughout subsequent interfaces.

For Ni/Ti:$^{11}B_4C$ multilayers the layers become more strongly correlated at higher lateral spatial frequencies, suggesting that short spatial details are more readily replicated in Ni/Ti:$^{11}B_4C$ multilayers than in pure Ni/Ti multilayers. Moreover, the addition of $^{11}B_4C$ interlayers to the Ni/Ti multilayers leads to an even stronger correlation at higher $q_y$ values, indicating highly correlated interfaces for these samples. It should be noted that a stronger correlation between the interfaces does not necessarily mean a higher interface width. Previous work has shown that a significant reduction in interface width can be obtained by the $^{11}B_4C$ co-deposition, while simultaneously having correlated and mounded interfaces [21]. Correlation between the interfaces can, however, be a relevant parameter for the growth of supermirrors where thousands of layers are needed to be deposited.

GISAXS characterization showed that $^{11}B_4C$ incorporation resulted in vertically correlated interfaces with interface mounds. However, X-ray and neutron reflectivity measurements indicated that the overall reflectivity performance of the $^{11}B_4C$-containing multilayers was significantly improved at the measured period of 48 Å. This suggests that while interface mounds were present due to $^{11}B_4C$ incorporation, their size was still smaller than the reduction in roughness. This is likely due to reduced intermixing, which explains the lower overall diffuse signal for pure Ni/Ti multilayers despite their higher interface width.

The low adatom mobility of $B_4C$ under the prevailing deposition conditions is well known [20], which results in an island-like growth, explaining the formation of mounded interfaces for the $^{11}B_4C$ interlayers. This behavior is consistent with existing theories on nucleation and multilayer growth [36,37]. Due to the shadowing effects, incoming adatoms arriving at an angle from a tilted magnetron source are blocked, resulting in faster growth of the mounds than the valleys. Consequently, a mounded interface is formed over time. While correlated and mounded interfaces do not necessarily imply a larger interface width [21], a strong correlation between the interface profiles may lead to roughness replication during multilayer growth, which can be important for the design of neutron supermirrors where thousands of layers need to be deposited.

To investigate the reflectivity performance of the various multilayer designs in different q regions, X-ray and neutron reflectivity measurements were conducted on the stacked multilayers. Figure 6 a) shows the neutron reflectivity of the five investigated multilayer stacks. The multilayer periods of 84 Å, 48 Å and 30 Å result in first-order Bragg peaks at diffraction angles of approximately $2\theta = 3.5°$, $6.5°$, and $10°$, respectively. Additionally, second-order Bragg peaks appear at around $2\theta = 7°$ and $11°$, while Kiessig interference fringes can be observed between the peaks.

In Figure 6 b) the first-order Bragg peaks for the 84 Å, 48 Å, and 30 Å multilayer periods are separated and plotted as a function of the scattering vector q. To compensate for the theoretical intensity, decrease for a flat interface according to Porod's law [38] and facilitate comparison of the Bragg peaks on a linear scale, the intensity has been scaled with $q^4$.

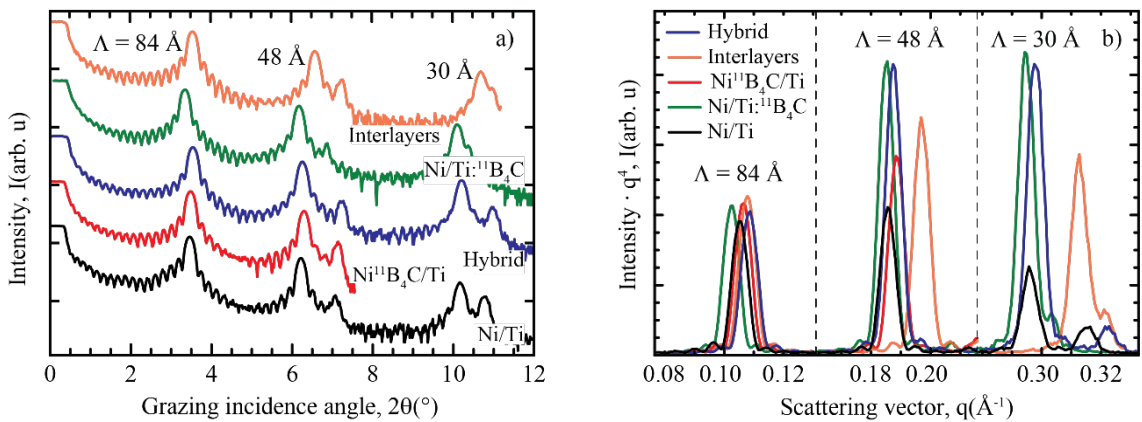

**Figure 6.** a) Neutron reflectivity of stacked multilayers with periodicities of 84 Å, 48 Å, and 30 Å, with different designs. b) First-order Bragg peaks for the different periodicities of the multilayers in a). The intensity is scaled by $q^4$ to aid visual comparison.

Analysis of the first order Bragg peaks for the 84 Å period revealed that the highest intensity is obtained for Ni/Ti multilayers with $^{11}B_4C$ interlayers, followed by $Ni^{11}B_4C$/Ti. This result is expected, as these are designed to have a large contrast in SLD, which is crucial in this q range, with added $^{11}B_4C$ to reduce the interface width. $Ni/Ti:^{11}B_4C$, which has the poorest SLD contrast among the investigated material designs but is known for its very low interface widths, has a lower intensity. Thus, here, a low interface width is not sufficient to compensate for a lack of scattering contrast. On the other hand, the Ni/Ti multilayer stack and the hybrid multilayer stack, where the 84 Å period multilayer consists of pure Ni/Ti, have the lowest intensities despite the highest SLD contrast, indicating large interface widths in these multilayers. Overall, it can be concluded that all $^{11}B_4C$-containing multilayers perform better than pure Ni/Ti for this period, with $^{11}B_4C$ interlayers exhibiting the best performance.

The trend is similar at the higher q ranges for periods of 48 Å and 30 Å. The $Ni/Ti:^{11}B_4C$ multilayers perform best, along with the hybrid multilayers consisting of $Ni/Ti:^{11}B_4C$ in the thinner periods. In these q ranges, where thinner multilayer periods are used, reflectivity is less sensitive to SLD contrast and more dependent on a low interface width. Adding $^{11}B_4C$ only to the Ni layers in $Ni^{11}B_4C$/Ti does not seem to reduce the interface width sufficiently compared to adding $^{11}B_4C$ in the entire multilayer, and the measured intensity is significantly lower than that of the $Ni/Ti:^{11}B_4C$ multilayer. The Ni/Ti multilayer has the lowest intensity, indicating a high interface width, which is likely why $^{11}B_4C$ interlayers do not perform well at this higher q range. Although the reflectivity measurement of the 30 Å period for $Ni^{11}B_4C$/Ti was interrupted, it is expected that its performance will not be better than that of $Ni/Ti:^{11}B_4C$. This is evident from neutron reflectivity measurements of $Ni^{11}B_4C$/Ti and $Ni/Ti:^{11}B_4C$ single multilayers with a period of 30 Å (not shown), where $Ni^{11}B_4C$/Ti shows a worse performance. Although not yielding the highest reflectivity in any q range, the hybrid multilayer design demonstrates relatively good performance across the entire q range. This highlights the potential of combining different material designs in multilayer structures to achieve maximum reflectivity over a broad range of q values.

To gain a better understanding of the reflectivity behavior, the structural parameters of the stacked multilayers were determined using coupled fitting of the X-ray and neutron reflectivity measurements. Figures 7 a) and b) show these fits, with the fitted reflectivity shown in red for X-ray and neutron reflectivity, respectively, for the best-performing multilayer designs based on the analysis of the first-order Bragg peaks shown in Figure 6. The Ni/Ti multilayer stack is also included for comparison.

Both $^{11}B_4C$-containing multilayers show good agreement with the coupled fits to the X-ray and neutron reflectivity data, indicating that the used model accurately describes their structure. However, a satisfying fit could not be obtained for the pure Ni/Ti multilayer stack, and a more sophisticated model is needed to describe this multilayer accurately. The observed discrepancy with the predicted simulations may be due to the accumulation of interface width and layer thicknesses throughout the multilayer, which has been observed in previous studies as well [3]. The clear broadening at the first Bragg peak in the X-ray reflectivity data indicates the presence of such behavior. Accumulating roughness can cause local variations in the multilayer period and effectively lower the number of multilayer periods contributing to the intensity, which could explain the observed X-ray reflectivity data. To assess the quality of the multilayer, the best fit to the first-order Bragg peaks in each period present in the stack was made. The results showed that an interface width of 10.5 Å on both the Ni and Ti layers is required for the reflectivity performance shown in both neutron and X-ray reflectivity. It should be noted that the fit to the neutron reflectivity is very good for the first four diffraction peaks.

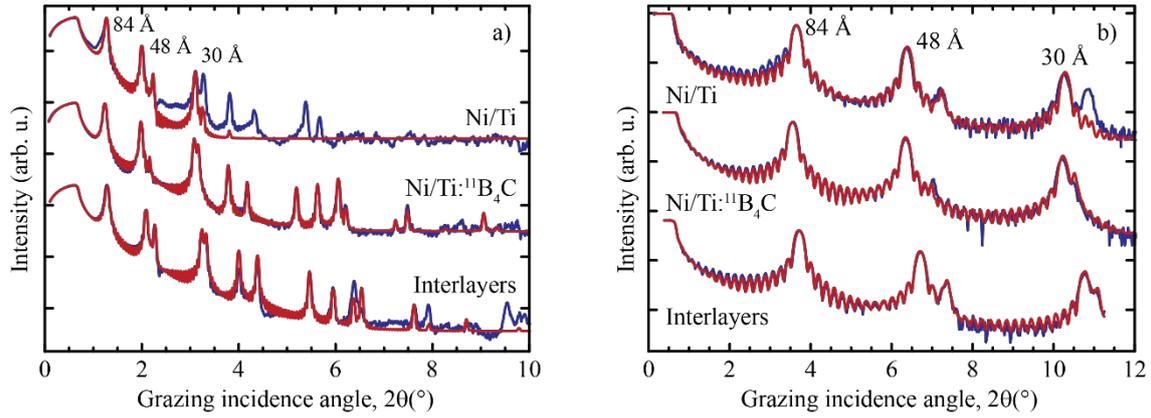

**Figure 7.** X-ray and neutron reflectivity measurements and fits for the Ni/Ti, Ni/Ti with $^{11}B_4C$ interlayers, and Ni/Ti:$^{11}B_4C$ multilayer stacks containing periods of 84 Å, 48 Å, and 30 Å. The fits were performed to both neutron and X-ray data simultaneously with coupled structural parameters using the GenX software. The measurements are vertically displaced for clarity.

The individual interface widths for the Ni and Ti layers were determined based on the reflectivity fits presented in Figure 7, as well as similar fits for X-ray and neutron reflectivity measurements of single multilayers for the various multilayer designs (not shown). It was found that the Ni interface was consistently wider than the Ti interface for all multilayers, as illustrated in Figure 8 a). This result is expected, as Ni tends to form crystallites more readily than Ti, as observed earlier and reported in the literature [17,18,19]. It can also be seen that the Ti interface width is about the same for all periods within each multilayer design.

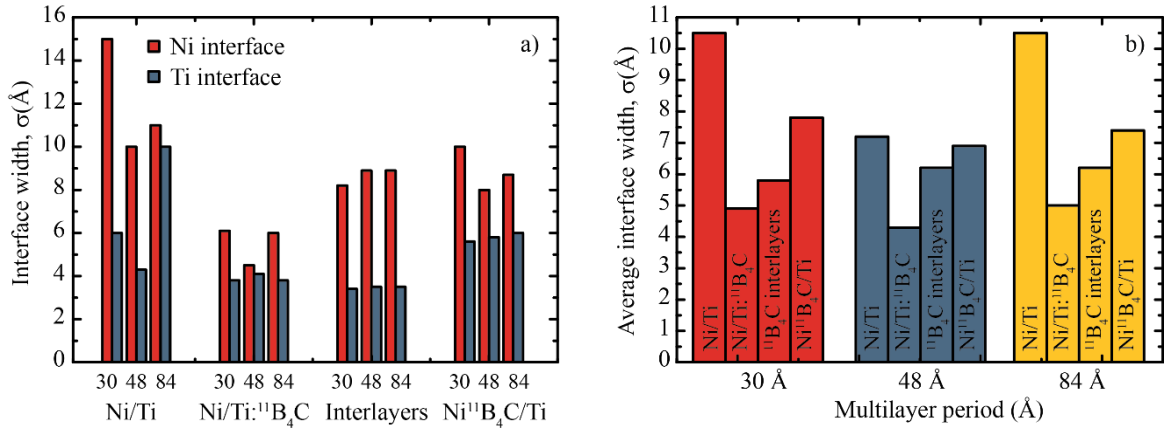

**Figure 8.** Interface widths obtained from coupled X-ray and neutron reflectivity fitting for the different multilayer designs and periods of 30 Å, 48 Å, and 84 Å. a) Individual interface widths for Ni and Ti. b) Average interface widths for each multilayer design.

However, it should be noted that determining individual interface widths from reflectivity fitting can be challenging. For neutron measurements, there are often not enough Bragg peaks available for thin period multilayers to accurately determine the individual interface widths. Additionally, similar figures of merits can sometimes be obtained if the fitted interface widths are swapped for the layers in the fit. Therefore, to increase the reliability of the results, average interface widths are commonly reported instead of individual ones. Figure 8 b) illustrates the average interface widths obtained in this study.

When comparing the average interface widths, it is observed that Ni/Ti has the highest interface width for all periods, with values up to 10.5 Å. This high value for the interface width may be attributed to the difficulties in obtaining a good fit for the X-ray and neutron reflectivity measurements simultaneously, as mentioned earlier. Therefore, this value might not be the actual physical interface width, although the multilayer has such a performance. The average interface widths of multilayer designs containing $^{11}B_4C$ remain constant regardless of the multilayer period. This is not the case for Ni/Ti where a low interface width was not achieved at 30 Å or 84 Å. At 30 Å, this is likely explained by the difficulty of forming a layered structure due to a compositional intermixing of Ni and Ti forming disordered phases at low thicknesses [2,], while at 84 Å, this might be due to roughness accumulation, which is not accounted for in the fitting model.

It can be noted that Ni/Ti:$^{11}B_4C$, which was found to be the optimal multilayer design for reflectivity in the high q region, has the lowest interface width of all designs for all periods, about 4.6 Å when averaged over all periods. Similarly, the Ni/Ti multilayers with $^{11}B_4C$ interlayers, which were the optimal multilayer design in the low q region, have a corresponding interface width of 6.2 Å. These two interface width values were used to predict the supermirror performance of a hybrid design.

*Material design for $^{11}B_4C$-containing Ni/Ti multilayer neutron supermirrors*

The results demonstrate the dependence of the optimal multilayer design on the multilayer period which means that neither a design optimized for low interface width, nor a design optimized for high optical contrast provide good reflectivity over the relevant q-range. The multilayer designs focused on maximizing optical contrast between layers, including Ni$^{11}$B$_4$C/Ti and $^{11}$B$_4$C interlayer multilayers, underperformed in the high q-region compared to the Ni/Ti:$^{11}$B$_4$C multilayer. However, these designs demonstrated the best reflectivity performance in the low q-region. Conversely, in the high q-region, the Ni/Ti:$^{11}$B$_4$C performed better than the designs optimized for high optical contrast.

Since different materials design are needed depending on the multilayer period this has clear implications for the development of neutron supermirrors with multiple thicknesses in the multilayer stack. In practice, a depth-graded approach is needed, where the multilayer design varies depending on the layer thickness in the stack. Figure 9 shows one such depth-graded material design according to this work's structural and optical properties findings. Since the best reflectivity at low q can be achieved for a Ni/Ti design with 11B4C interlayers and from a scattering vector of $q_z = 0.11$ Å-1, corresponding to a multilayer period of 52 Å and higher, the Ni/Ti:$^{11}$B$_4$C material system is expected to perform better. By applying the optimal material system in either region in q-space, a supermirror can be constructed that shows a higher reflectivity than a state-of-the-art commercially available Ni/Ti multilayer with an interface width of 7 Å.

Furthermore, as it is demonstrated that the Ni/Ti:$^{11}$B$_4$C material system can be grown successfully for thin periods of 30 Å with similar interface widths, this material system can also be used for high-m supermirrors, although at the expense of more required layers in the supermirror where the total amount of required layers roughly scales with $N = 4m^4$ [39]. The m = 7 Ni/Ti supermirror in Figure 9 has been simulated with a total of 5000 layers with layer thicknesses ranging from 20 Å to 400 Å. The m = 10 hybrid mirror where two material designs were combined is simulated using N = 15000 layers ranging from 15 to 400 Å in layer thickness. The total thickness variation in the simulation follows a power law similar to that introduced in 1977 by Mezei et al [40,41],; while there exist more effective algorithms for neutron supermirrors [42,43,44], this simple approach is still useful to demonstrate the relative performance of different multilayer designs.

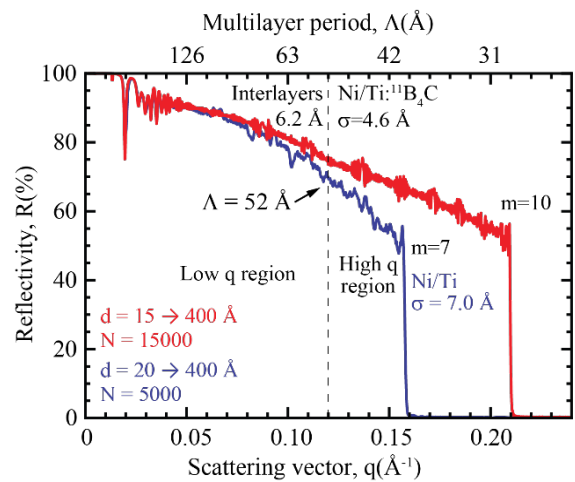

**Figure 9.** The calculated performance of a hybrid supermirror design compared to that of a conventional Ni/Ti supermirror. The blue curve shows a conventional Ni/Ti multilayer with an interface width of σ = 7.0 Å. The red curve shows the simulated reflectivity from a hybrid multilayer design with $^{11}$B$_4$C interlayers with an interface width of σ = 6.2 Å for layers thicker than 26 Å (corresponding to a multilayer period of 52 Å), and Ni/Ti:$^{11}$B$_4$C with an interface width of σ = 4.2 Å for layers thinner than 26 Å.

**Conclusion**

This study investigated different materials designs using single materials combination of Ni, Ti, and $^{11}$B$_4$C targets during ion assisted magnetron sputter deposition for depth-graded multilayers such as supermirrors for neutron optics. Combined structural characterization and neutron reflectivity simulations revealed that the optimal material system for a given neutron multilayer depends on the period used for the multilayer stack. Strategies that focus on the optimization of the achieved optical contrast show very good results for thicker periods, but for shorter periods, the interface width becomes more important. Therefore, several material systems must be combined to maximize the reflectivity of a supermirror over a large scattering vector q-range. Neutron reflectivity measurements show that at a multilayer period of 84 Å, the best reflectivity performance is obtained for Ni/Ti multilayers where thin $^{11}$B$_4$C interlayers were deposited at the interfaces, while for shorter periods of 48 Å and 30 Å, a Ni/Ti:$^{11}$B$_4$C multilayer where $^{11}$B$_4$C has been co-deposited throughout the entire multilayer stack should be used. For depth-graded multilayers such as supermirrors, this means that multiple material systems need to be combined to maximize reflectivity. The optimal neutron supermirror proposed in this work would, therefore, consist of Ni/Ti layers with $^{11}$B$_4$C interlayers at the interfaces for layer thicknesses above d = 26 Å, while thinner layers are grown with $^{11}$B$_4$C co-deposited throughout the entire stack. Supermirror simulations show that a clear improvement in neutron reflectivity can be obtained compared to conventional Ni/Ti supermirrors, which would result in a higher neutron flux being available for experiments. As the achieved flux is important for science conducted with neutron scattering, such an improvement would have a large and immediate impact on conducted neutron experiments.

## Author Contributions

**Sjoerd Stendahl:** conceptualization, methodology, synthesis, formal analysis of X-ray/ neutron and Synchrotron experiments.
**Naureen Ghafoor:** conceptualization, methodology, neutron and Synchrotron experiments, TEM analysis.
**Anton Zubayer:** Assistance in synchrotron experiments.
**Marcus Lorentzon:** Assistance in the deposition process.
**Alexei Vorobiev:** Assistance in neutron scattering experiments.
**Jens Birch:** conceptualization, supervision, funding acquisition.
**Fredrik Eriksson:** supervision, project administration, conceptualization, methodology, neutron and synchrotron experiments.
**SS** and **NG** wrote the first draft and all co-authors participated in writing, reviewing, and editing.

## Acknowledgments

This work was supported by the Swedish Foundation for Strategic Research (SSF) and the Swedish national graduate school in neutron scattering, SwedNess (grant number: GSn15-0008), and the Swedish Research Council (VR). JB acknowledges the Swedish Government Strategic Research Area in Materials Science on Advanced Functional Materials at Linköping University (Faculty Grant SFO Mat LiU No. 2009 00971) for financial support. The authors would like to acknowledge the support from the Swedish Research Council VR-RFI (#2017-00646_9) for the Accelerator based ion-technological center, and from the Swedish Foundation for Strategic Research (contract RIF14-0053) for the tandem accelerator laboratory in Uppsala for assistance with elastic recoil detection analysis.

## Data Availability

The raw data required to reproduce these findings cannot be shared at this time due to technical or time limitations but will be available to download after the review process.